# Patient Independent Interictal Epileptiform Discharge Detection

Matthew McDougall[1,*], Hezam Albaqami[1,2], Ghulam Mubashar Hassan[1], and Amitava Datta[1]

*Abstract—* **Epilepsy is a highly prevalent brain condition with many serious complications arising from it. The majority of patients which present to a clinic and undergo electroencephalogram (EEG) monitoring would be unlikely to experience seizures during the examination period, thus the presence of interictal epileptiform discharges (IEDs) become effective markers for the diagnosis of epilepsy. Furthermore, IED shapes and patterns are highly variable across individuals, yet trained experts are still able to identify them through EEG recordings – meaning that commonalities exist across IEDs that an algorithm can be trained on to detect and generalise to the larger population. This research proposes an IED detection system for the binary classification of epilepsy using scalp EEG recordings. The proposed system features an ensemble based deep learning method to boost the performance of a residual convolutional neural network, and a bidirectional long short-term memory network. This is implemented using raw EEG data, sourced from Temple University Hospital's EEG Epilepsy Corpus, and is found to outperform the current state of the art model for IED detection across the same dataset. The achieved accuracy and Area Under Curve (AUC) of 94.92% and 97.45% demonstrates the effectiveness of an ensemble method, and that IED detection can be achieved with high performance using raw scalp EEG data, thus showing promise for the proposed approach in clinical settings.**

## I. Introduction

Epilepsy affects more than 1% of the world's population and is a brain condition that causes repeated seizures – moments of uncontrolled electrical activity within the brain causing abnormal behaviours, sensations, or movements, and loss of consciousness [1]. If left untreated this poses many health consequences such as shortened life span, bodily injury, neuropsychological impairments and social disabilities [2]. Seizures, however, are not distinct markers of epilepsy as many factors are involved in causing one to occur. Interictal epileptiform discharges (IEDs) are specific markers of epilepsy, characterized by sharp, spikey waves that are temporary in nature and are not continuous or uninterrupted like seizure activity [3]. They are useful diagnostic observations due to their localisation and frequency [3,4].

Electroencephalography (EEG) is the standard tool used to diagnose seizure and epilepsy. EEG measures the brain activities and a physician manually analyzes the recordings for significant markers. The process is time-consuming, especially with long-term EEG recordings. Therefore, the need to automate or assist such a process using artificial intelligence is growing.

With the ever-increasing access to more data, there is a growing interest in utilising deep learning technologies for data analysis [4]. The most commonly implemented networks are Convolutional Neural Networks (CNN, 1D and 2D) [5], Recurrent Neural Networks (RNN) [6], Auto-Encoders (AE) [7], Deep Belief Networks (DBN) and Multi-Layer Perceptron (MLP) [8]. CNNs are proving to be powerful tools in deep learning due to their computational efficiency and ability to extract features with high inter-class variance [9, 10, 11]. Similarly, RNNs are useful for sequence and temporal series processing due to their ability to retain memory, making them particularly useful for processing long sequences of data [9]. Moreover, ensemble methods [12] allows to combine the predictions made from multiple classifiers as a way to cancel out bias from each other [9]. Usman et al. [10] showed that for their study of seizure detection, the addition of an ensemble classifier boosted the accuracy from 92.98% to 94.31%.

Many research studies developed for patient-specific data achieve high levels of performance. However, the similar distribution of training and testing data in this approach allows for better prediction accuracy [5,13]. IEDs are very physiologically dependent events, and thus their morphology may vary a lot across individuals which makes detection of specific signals difficult [14].

Few recent studies investigate the patient independent detection of IEDs as shown in **Table 1**. However, the majority of these studies utilised datasets which contain annotated IEDs thus allowing them to deliver targeted information to their models. For the studies mentioned in **Table 1**, only two of the datasets are easily accessible – TUEP and TUEV. Furthermore, the number of subjects in previous research vary depending on the dataset used, where ideally a larger amount of subjects would be preferred so the patient independent performance can be more plausible. Overall, there is a diverse range of methods used in the detection of IEDs which shows the many possibilities that researchers can use, but also makes it difficult to identify which models are truly the best among them all.

This study proposes a patient independent deep learning ensemble model for the binary classification task of diagnosing epilepsy. This model is based on residual CNN and bidirectional Long short term (Bi-LSTM) sub models, and aims to generalise the learning of IED patterns across individuals. The motivation for this work comes from the need to develop simpler, computationally inexpensive models, that can be more easily compared to other studies, which is done by using a widely accessible dataset and raw scalp EEG data.

[1] Department of Computer Science and Software Engineering, The University of Western Australia, Australia. * Corresponding Authors: 22494429@student.uwa.edu.au, haalbaqamii@uj.edu.sa (M. McDougall; H. Albaqami).

[2] Department of Computer Science and Artificial Intelligence, The University of Jeddah, Saudi Arabia.

Table 1: State of the art methods for patient independent IED detection

| Reference | Model | Dataset | Features | #Subjects | #IEDs | Performance |
|---|---|---|---|---|---|---|
| **Thomas et al. [15]** | CNN + SVM | EEG, MGH | Temporal | 93 | 14 364 | 94% AUC, 83.86% ACC |
| **Golmohammadi et al. [14]** | Autoencoder | EEG, TUEV | Temporal + Spatial | 390 | 19 057 | 90.1% SEN, 95.1% SPE |
| **Uyttenhove et al. [16]** | CNN | EEG, TUEP | Temporal | 200 | 123 815 | 93.2% AUC, 81.4% ACC |
| **Sabor et al. [17]** | LSTM | EEG, TUEV | Frequency | 390 | 19 057 | 95.5% ACC, 95.4% SEN |
| **Geng et al. [18]** | IEDnet | iEEG, Independent | Temporal | 7 + 5 | 4034 + 2078 | > 93% AUC, > 93% ACC |

AUC: area under curve, ACC: accuracy, SEN: sensitivity, SPE: specificity, iEEG: intracranial EEG

## II. METHODOLOGY

### A. Dataset and Pre-processing

Our study is based on the Temple University Hospital EEG Epilepsy Corpus (TUEP), (see **Table 2**), [19] which contains scalp EEG data on 100 subjects with epilepsy and 100 subjects without epilepsy. This dataset consists of 1360 files with epilepsy and 288 files without epilepsy. The EEG recordings were all sampled at 250 Hz and vary greatly in length for each session – from tens of seconds to hours; they also vary in the number of channels used – from 24 to 33; and lastly they vary in the referential montages used – being either Linked Ear or Averaged Referenced, although both still utilise the 10-20 standard montage. Most of the epileptic patient recordings were taken during the interictal periods, and several sessions captured seizure events.

Out of the 1360 files from the epileptic patients, only 623 files contained relevant information i.e. IEDs. The session reports of the remaining files noted that there were no epileptiform discharges present in the recording or that it appeared normal, thus these files were removed.

To maintain consistency across all recordings, 30 channels were taken, and for those with less than 30 channels, the first $n$ channels were recycled where $n$ is the difference between 30 and the number of good channels of the recording. The channels were chosen to be oversampled to 30 rather than under sampled to a smaller amount to avoid loss of information where some channels may be particularly significant. Furthermore, there was too great a difference in the length of recordings to handpick segments from each, thus for simplicity, the first 30 seconds of each was used. The TUEP dataset contained segments that were cut from their original recordings, so that only significant recordings were made available. Thus, the decision to use the first 30 seconds was done under the assumption that they were cut to the time which significant events begun to occur. Where recordings were less than 30 seconds in duration, the remaining length was padded with zeros. EEG signal frequencies range from 0.5 to 128 Hz [16] and power line frequencies exist at 50 or 60 Hz [4] therefore a second order Butterworth bandpass filter between 0.5 and 49 Hz was applied to the signals to remove these artefacts. Finally, the data was normalised channel-wise using the StandardScaler function within the scikit-learn library in python.

### B. Proposed Model

The proposed model is a combination of two sub models, through an ensemble classifier which brings everything together as one. Each sub model, and the ensemble, are compiled with a binary cross-entropy loss function and RMSprop optimizer (see **Figure 1**).

#### 1) Residual Convolutional Neural Network (CNN)

A customised ResNet is built which consists of three main convolutional batches. All convolution layers are applied with 128 filters, kernel size of 5, with a tanh activation function. For batches 1 and 2, they contain two convolutions in total which are each followed by a max pooling step with pool size of 5, and a dropout layer with a rate of 0.5 to combat overfitting during training. Batch 3 is slightly different as it instead contains three convolutional layers which are each followed by max pooling steps with pool sizes of 3, 2, and 2 respectively. No dropout is used in the third batch. The final layer of batches 2 and 3 up sample the data to match the same dimensions so that the different levels of features learnt after each batch can then be summed together at the end. Global average pooling is then performed. Two fully connected layers with tanh activation and dropout of 0.5 in between, are used to perform the classification of the signal. The final out put dense layer is applied with the sigmoid activation function. The total number of trainable parameters for the sub model is 536,449.

#### 2) Bi-directional Long Short-Term Memory (Bi-LSTM)

A convolutional layer with 32 filters, kernel size of 5, and ReLU activation is first applied to the input data in order to reduce the computational cost of using the Bi-LSTMs. This is followed by a max pooling step with a pool size of 5, and a dropout of 0.5. Two Bi-LSTMs are used in succession, with 64 and 32 units respectively, each followed by a ReLU activation function and dropout of 0.5. Each of the Bi-LSTM layers have a sequence-to-sequence architecture. Global average pooling is performed and fed into two fully connected layers with tanh activation, to perform the classification. The final output dense layer is applied with the sigmoid activation function. The total number of trainable parameters for the sub model is 62,945.

Table 2: Summary of data used for training and testing

| Diagnosis | | No. Patients | No. Files |
|---|---|---|---|
| Epileptic | Training | 32 | 451 |
| | Testing | 10 | 172 |
| Non-Epileptic | Training | 80 | 224 |
| | Testing | 20 | 64 |
| Total | | 142 | 911 |

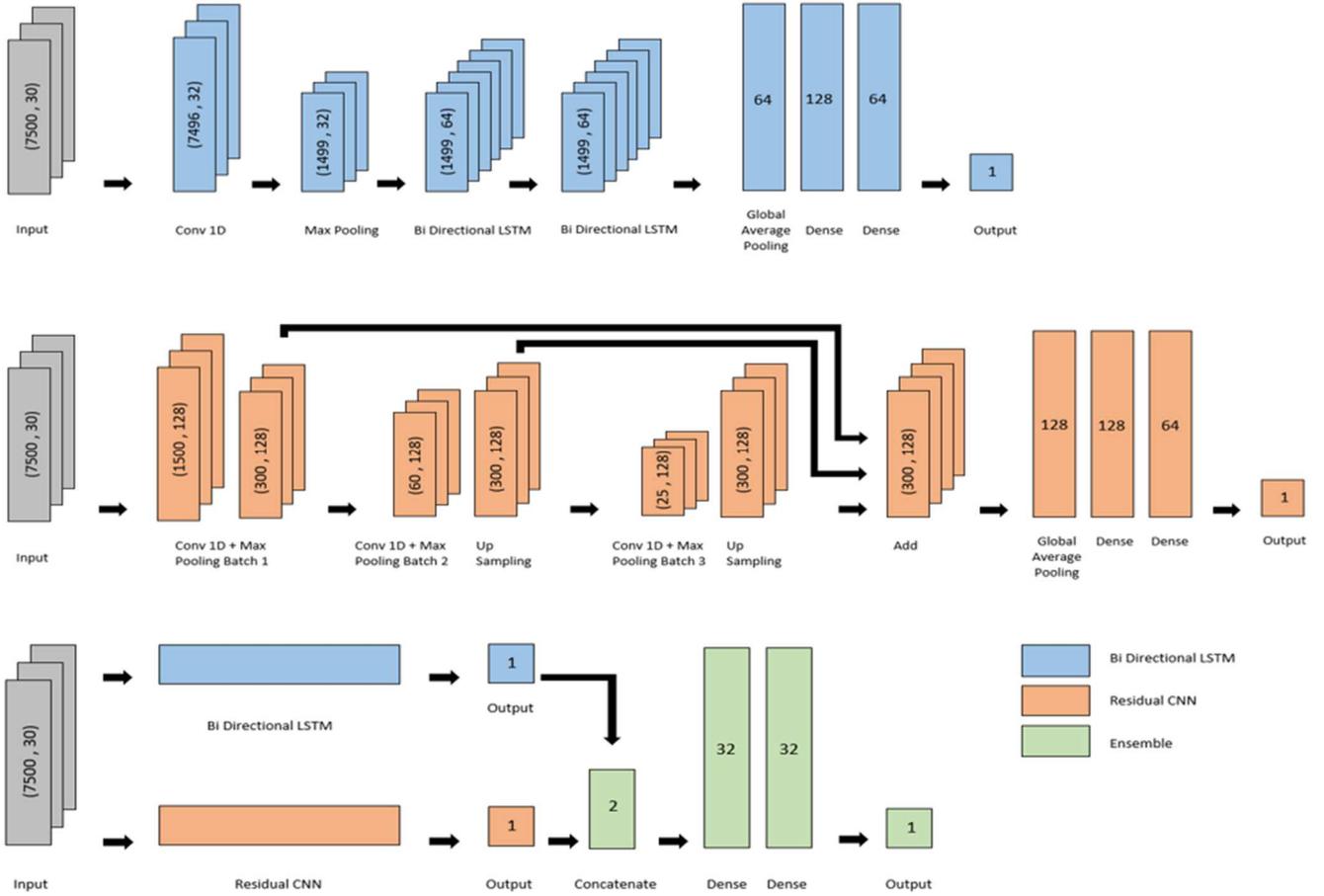

**Figure 1**: Architecture of bidirectional LSTM (top) and residual CNN (middle) sub models, and ensemble (bottom)

*3) Ensemble Classifier*

The fully trained Bi- LSTM and residual CNN are imported with their layers frozen to prevent further training. The single probability outputs from each are concatenated and fed into two fully connected layers with tanh activation, and then a single output dense layer with sigmoid activation. The total number of parameters for the ensemble is 600,579 however since both of the sub models are frozen, the only layers being trained are the fully connected layers. The number of trainable parameters therefore is 1,185.

*C. Training and Testing*

To evaluate the proposed model on its ability to generalise across patients, the dataset was split so that no single patient had recordings present in both the training and testing set. This way, the model is being tested on data from patients it has not seen before, and a true evaluation of its generalisation ability can be made. An approximate 20% of the patients from each of the epileptic and non-epileptic populations were used for testing. The training data was split further for training 80% and validation 20%. The current imbalances in the training and validation data between epileptic and non-epileptic files would bias the model and hinder its performance – where the model can simply learn to predict every recording as epileptic and still manage to achieve moderately high performance. To avoid this, and to prevent loss of information, the minority

**Table 3:** Summary of proposed intra-model performance and comparison to the current state of the art

| Reference | Model | AUC | Acc | Sens | Spec | F1 |
|---|---|---|---|---|---|---|
| **Uyttenhove et al. [16]** | CNN | 0.9552 | 0.7650 | 0.7589 | 0.7857 | -- |
| | CNN GAP | 0.9315 | 0.8142 | 0.8156 | 0.8095 | -- |
| **Current Study** | Residual CNN | 0.9529 | 0.9280 | 0.9360 | 0.9062 | 0.9499 |
| | Bi-directional LSTM | 0.9652 | 0.9195 | 0.9186 | **0.9219** | 0.9433 |
| | Ensemble | **0.9745** | **0.9492** | **0.9651** | 0.9062 | **0.9651** |

class – non-epileptic, was oversampled to twice the amount to balance with the number of epileptic files. Both sub models, and the ensemble were trained individually with the validation loss used as a monitor to identify when overfitting occurs, and to tune the hyper-parameters. Training ceased when loss was no longer improving.

The proposed method was implemented within Python 3.9.7, using Keras and TensorFlow open-source libraries. Processing is performed with a system operating on a Ryzen-5 3400G central processing unit, Nvidia 2060 super with 12GB of video memory, and 16GB of random-access memory.

## III. RESULTS AND DISCUSSION

A summary of the results is given in **Table 3** where a comparison against the current state of the art model is also made. **Figure 2** tracks the training and validation loss of both the Bi-LSTM and residual CNN sub models, showing that it did not require many epochs for each model until they begun to overfit the data. The optimal version of each sub model was chosen from the epoch before it begun to overfit. Both sub models required no longer than approximately 20 minutes until training was completed. Furthermore, since the classifier within the ensemble was only working with the small concatenation tensor of two values, optimal performance and training of the model was complete within one epoch.

### A. Intra-Model Performance Evaluation

The ensemble method demonstrates improvements for all metrics except for the specificity. Through the concatenation of the single output probabilities from both the Bi-LSTM and residual CNN, the ensemble was able to learn the best possible method of combination to achieve greater results than that from each of the two individually. There appears to be no largely dominant sub model when comparing the bidirectional LSTM and the residual CNN, as one outperforms the other for the area under curve (AUC) and specificity, whilst the other performs better for accuracy, sensitivity, and F1. Therefore both sub models significantly contributed information to the learning of the final ensemble model, albeit not entirely equally, however.

From the confusion matrix in **Table 4**, the residual CNN is able to correctly predict a total of three more epileptic recordings than the Bi-LSTM, at the cost of incorrectly predicting one more non-epileptic recording. The ensemble greatly improves on this by correctly predicting five more than the residual CNN, thus increasing the sensitivity to a clinically acceptable level [20, 21]. Through **Figure 3** we see how the ensemble appears to have more predictions in common with the residual CNN than it does with the Bi-LSTM, regardless of whether they were correct or incorrect. This means that the ensemble favoured the predictions made by the residual CNN for the majority of the cases, perhaps because it was more likely to be correct than the other one, which affected the ensemble's decisions for the incorrect classifications. Despite this, there were four instances of correct classification where the ensemble favoured the decision of the Bi-LSTM, showing that it wasn't simply leaching off of one sub model, and knew how to leverage the information from both. Finally, it is worth noting that there was never a case where the ensemble went against the decisions of both sub models and misclassified the recording, but there was one case where it overturned their decisions to yield a correct classification of an epileptic signal. When both sub models are confident in their decision of it being epileptic or not, they generally express values ≥ 0.8 and ≤ 0.2 and the ensemble has no reason to go against them. For that single epileptic recording (**Figure 4**), however, the sub models both expressed less confident values which allowed the ensemble to overturn their predictions. Thus the combined degree of confidence of both sub models also influences the ensemble's decision.

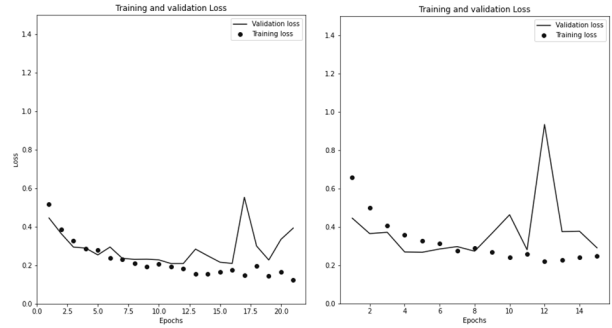

**Figure 2**: Training and validation loss for bidirectional LSTM (left) and residual CNN (right).

**Table 5** shows the corresponding mean probabilities that each sub model, and the ensemble give when they decide on a recording as being epileptic or not – noting that this does not exactly mean that their predictions were correct. Whilst the ensemble manages to achieve greater accuracy overall, we see here that it yields the least confident prediction out of them all, with the residual CNN being the most confident and the Bi-LSTM coming in between. Naturally this is what we would expect of an ensemble that has to constantly weigh the decisions of both sub models whilst still aiming to achieve a greater accuracy overall. Even though it is the least confident, it still has an AUC value of 97.45% meaning that it has a strong distinction between epileptic and non-epileptic classes. **Figure 4** demonstrates examples of how the ensemble would on occasions favour the decision of the Bi-LSTM sub model, whilst at other times favouring the decision of the residual CNN. The difference in output probabilities for each of the sub models may be dependent on the features extracted from each, perhaps misidentifying spikes or confusing them with artefacts. To fully understand the reasoning behind their interpretation, professional input is needed to understand what events are present in each of these signals. Ensembles work best when their constituents perform well whilst also being as different as possible [9]. When both models perform well they come to a consensus on the majority of recordings, as in the case here where both sub models achieved high accuracies and correctly predicted the same recordings (**Figure 3**). There is little learning and decision making the ensemble needs to do for these cases. The ensemble, therefore, would be able to at least achieve the same level of accuracy because it could simply favour one of the sub models and replicate their results. The true learning, and subsequently improvement in results, comes through the diversity of the sub models, which causes

Table 4: Confusion matrix for both sub models and the ensemble.

|  |  | Residual CNN | | Bi-LSTM | | Ensemble | |
|---|---|---|---|---|---|---|---|
|  |  | NE | E | NE | E | NE | E |
| Ground Truth | NE | 58 | 6 | 59 | 5 | 58 | 6 |
|  | E | 11 | 161 | 14 | 158 | 6 | 166 |

E: Epileptic, NE: Non-Epileptic

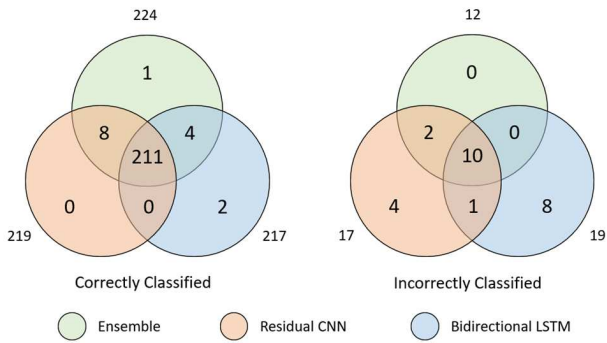

Figure 3: Venn diagram of correct and incorrect classifications of both sub models and the ensemble.

Table 5: Mean (95% Confidence Interval) prediction probabilities for both CNN, Bi-LSTM and ensemble models.

| Predication | Residual CNN | Bi-LSTM | Ensemble |
|---|---|---|---|
| Epileptic | 0.982 (0.866-1) | 0.905 (0.866-1) | 0.891 (0.764-1) |
| Non-Epileptic | 0.016 (0.-0.07) | 0.174 (0.-0.41) | 0.34 (0.26-0.42) |

their predictions to vary from each other. This is the essence of how ensembles work, where differing opinions allow it to be more informed to make a better decision. The ensemble has to then learn the best combination of the two in order to correctly predict the outcome. The lack of improvement for specificity may be attributed to the fact that both sub models learnt to detect the same features within the non-epileptic recordings, and thus they agreed on the same decision for most of them. Within the training dataset, there were fourteen instances, or potentially less considering the non-epileptic recordings were doubled, in which the Bi-LSTM and residual CNN did not agree on. For the epileptic recordings, however, there were twenty-seven instances where the two sub models did not agree on, which could be due to each of them learning slightly different representations of the signal. The increase in disagreement between the Bi-LSTM and the residual CNN forced the ensemble to learn which of the two made a better decision for the recording in each situation, and to then favour its decision more. Carrying this over to the testing dataset, we noticed the same pattern where both sub models disagreed on three cases for the non-epileptic recordings, and eleven cases for the epileptic ones. Since the ensemble was able to learn more through the epileptic recordings on the training dataset, it was able to expand upon the results and to turn more of the false negatives from each sub model, into true positives, therefore increasing the sensitivity. The lack of disagreement between the sub models for the non-epileptic recordings combined with the ensemble's limited opportunity to learn from their predictions from the training dataset, prevented it from improving upon the specificity performance.

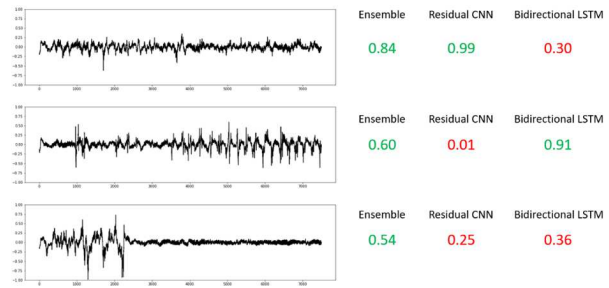

Figure 4: Epileptic recordings from three different patients in the testing dataset, and the assigned probabilities from each sub model and the ensemble. Note: the signals displayed are taken from one channel within the full 30 channel array, and the probabilities were based off the model assessing the entire 30 channel, 30 second recording.

### B. Comparison with State of the Art

To further validate the performance of the proposed model, a comparison was made against the current state of the art model by Uyttenhove et al. [16] where they utilised the same TUEP dataset, and the same aim of epilepsy diagnosis through detection of IEDs. They utilised two different CNN architectures – the Tiny-Visual Geometry Group and its global average pooling variant. Our proposed model was able to outperform their model for all performance metrics, achieving AUC, accuracy, sensitivity, and specificity of 97.45%, 94.92%, 96.51% and 90.62% respectively.

### IV. CONCLUSION

The work in this paper is able to demonstrate a successful implementation of a deep learning ensemble-based model for the detection of interictal epileptiform discharges (IEDs), to aid the diagnosis of epilepsy. It has learnt to identify the commonalities of these IEDs that exist across patients, making it a generalised, patient-independent model. These qualities highlight its potential for application within the clinical setting, as an additional diagnosis tool besides the examiner. This paper thus highlights the use of deep learning for epilepsy diagnosis beyond just the detection of seizures, and shows promise for the development of simple and generalised models that are more likely to be applicable within clinical settings. In future, we plan to investigate the use of different representations of the EEG data during the learning process, for more generalization capability.

### REFERENCES

[1] H. Albaqami, G. M. Hassan, A. Subasi, and A. Datta, "Automatic detection of abnormal EEG signals using wavelet feature extraction and gradient boosting decision tree," Biomedical Signal Processing and Control, vol. 70, p. 102957, 2021.

[2] H. Albaqami, G. M. Hassan, and A. Datta, " Wavelet-Based Multi-Class Seizure Type Classification System," Applied Sciences, 12 (2022) 5702.

[3] J. W. Britton *et al.*, "Electroencephalography (EEG): an introductory text and atlas of normal and abnormal findings in adults, children, and infants," 2016.

[4] S. Saminu *et al.*, "A Recent Investigation on Detection and Classification of Epileptic Seizure Techniques Using EEG Signal," *Brain Sciences,* vol. 11, no. 5, p. 668, 2021.

[5] X. Zhang, L. Yao, M. Dong, Z. Liu, Y. Zhang, and Y. Li, "Adversarial representation learning for robust patient-independent epileptic seizure detection," *IEEE journal of biomedical and health informatics,* vol. 24, no. 10, pp. 2852-2859, 2020.

[6] H. Daoud and M. A. Bayoumi, "Efficient epileptic seizure prediction based on deep learning," *IEEE transactions on biomedical circuits and systems,* vol. 13, no. 5, pp. 804-813, 2019.

[7] M. Golmohammadi, A. H. Harati Nejad Torbati, S. Lopez de Diego, I. Obeid, and J. Picone, "Automatic analysis of EEGs using big data and hybrid deep learning architectures," *Frontiers in human neuroscience,* vol. 13, p. 76, 2019.

[8] M. S. N. Chowdhury, A. Dutta, M. K. Robison, C. Blais, G. A. Brewer, and D. W. Bliss, "Deep neural network for visual stimulus-based reaction time estimation using the periodogram of single-trial eeg," *Sensors,* vol. 20, no. 21, p. 6090, 2020.

[9] F. Chollet, *Deep learning with Python*. Simon and Schuster, 2021.

[10] S. M. Usman, S. Khalid, and M. H. Aslam, "Epileptic seizures prediction using deep learning techniques," *Ieee Access,* vol. 8, pp. 39998-40007, 2020.

[11] R. Jana and I. Mukherjee, "Deep learning based efficient epileptic seizure prediction with EEG channel optimization," *Biomedical Signal Processing and Control,* vol. 68, p. 102767, 2021.

[12] S. Muhammad Usman, S. Khalid, and S. Bashir, "A deep learning based ensemble learning method for epileptic seizure prediction," *Computers in Biology and Medicine,* vol. 136, p. 104710, 2021/09/01/ 2021, doi: https://doi.org/10.1016/j.compbiomed.2021.104710.

[13] S. Elgohary, S. Eldawlatly, and M. I. Khalil, "Epileptic seizure prediction using zero-crossings analysis of EEG wavelet detail coefficients," in *2016 IEEE conference on computational intelligence in bioinformatics and computational biology (CIBCB)*, 2016: IEEE, pp. 1-6.

[14] Y. Roy, H. Banville, I. Albuquerque, A. Gramfort, T. H. Falk, and J. Faubert, "Deep learning-based electroencephalography analysis: a systematic review," *Journal of neural engineering,* vol. 16, no. 5, p. 051001, 2019.

[15] J. Thomas, L. Comoretto, J. Jin, J. Dauwels, S. S. Cash, and M. B. Westover, "EEG classification via convolutional neural network-based interictal epileptiform event detection," in *2018 40th Annual International Conference of the IEEE Engineering in Medicine and Biology Society (EMBC)*, 2018: IEEE, pp. 3148-3151.

[16] T. Uyttenhove, A. Maes, T. Van Steenkiste, D. Deschrijver, and T. Dhaene, "Interpretable epilepsy detection in routine, interictal eeg data using deep learning," in *Machine Learning for Health*, 2020: PMLR, pp. 355-366.

[17] N. Sabor, Y. Li, Z. Zhang, Y. Pu, G. Wang, and Y. Lian, "Detection of the interictal epileptic discharges based on wavelet bispectrum interaction and recurrent neural network," *Science China Information Sciences,* vol. 64, no. 6, pp. 1-19, 2021.

[18] D. Geng, A. Alkhachroum, M. A. M. Bicchi, J. R. Jagid, I. Cajigas, and Z. S. Chen, "Deep learning for robust detection of interictal epileptiform discharges," *Journal of neural engineering,* vol. 18, no. 5, p. 056015, 2021.

[19] L. Veloso, J. McHugh, E. von Weltin, S. Lopez, I. Obeid, and J. Picone, "Big data resources for EEGs: enabling deep learning research," in *2017 IEEE Signal Processing in Medicine and Biology Symposium (SPMB)*, 2017: IEEE, pp. 1-3.

[20] H. Albaqami, G. M. Hassan, A. Datta, "MP-SeizNet: A multi-path CNN Bi-LSTM Network for seizure-type classification using EEG," *Biomedical Signal Processing and Control*, Vol. 84, p. 104780, 2023.

[21] D. Hu, J. Cao, X. Lai, Y. Wang, S. Wang, and Y. Ding, "Epileptic state classification by fusing hand-crafted and deep learning EEG features," *IEEE Transactions on Circuits and Systems II: Express Briefs,* vol. 68, no. 4, pp. 1542-1546, 2020.